\documentclass[letter,12pt]{article}
\usepackage{graphicx,amssymb,amsmath,color,float}
\usepackage{jheppub} 
\usepackage[T1]{fontenc}

\def\beq{\begin{equation}}
\def\eeq{\end{equation}}

\def\bea{\begin{eqnarray}}
\def\eea{\end{eqnarray}}
\def\bei{\begin{itemize}}
\def\eei{\end{itemize}}
\def\bmat{\begin{matrix}}
\def\emat{\end{matrix}}
\def\={\,=\,}
\def\+{\,+\,}
\def\-{\,-\,}
\def\GeV{\,{\rm GeV}\,}

\def\MET{E_T^{\textrm{miss}} }
\def\mSFOS{m{\rm SFOS}}

\newcommand{\Fig}[1]{Fig.~\ref{#1}}
\newcommand{\Eq}[1]{eq.(\ref{#1})}

\preprint{EFI-13-15, ANL-HEP-PR-13-35, KIAS-P13040}

\author[a,b]{Stefania Gori}
\author[a,c]{Sunghoon Jung}
\author[a,d]{Lian-Tao Wang}

\affiliation[a]{Enrico Fermi Institute and Department of Physics, University of Chicago, Chicago, IL,
60637}
\affiliation[b]{HEP Division, Argonne National Laboratory, 9700 Cass Ave., Argonne, IL 60439}
\affiliation[c]{School of Physics, Korea Institute for Advanced Study, Seoul 130-722, Korea}
\affiliation[d]{Kavli Institute for Cosmological Physics, University of Chicago, Chicago, IL, 60637}

\emailAdd{goris@uchicago.edu}
\emailAdd{nejsh21@kias.re.kr}
\emailAdd{liantaow@uchicago.edu}

\title{Cornering electroweakinos at the LHC}

\abstract{Squeezed supersymmetric spectra are challenging for the LHC searches based on a sizable missing energy and hard visible particles. One such scenario consists of  chargino/second-lightest neutralino NLSPs and a lightest neutralino LSP with a relatively small mass gap ($m(\chi_{{\rm{NLSP}}})-m(\chi_{{\rm{LSP}}})\equiv\Delta\sim(10-50)$ GeV). 
In this note, we explore search strategies to better probe this parameter space. We focus on the $3\ell$ + $\MET$ channel arising from the chargino/second-lightest neutralino associated production, and we investigate the role of a relatively hard initial state radiation (ISR) jet. In addition to typical kinematic variables, such as the minimum lepton pair invariant mass, we  propose an angular separation variable and two ratio variables which capitalize on the main kinematic features; leptons stay relatively soft under the boost from ISR and a sizable missing energy arises only in tight correlation with the ISR boost. With 300/fb of data at the 14 TeV LHC, we expect to probe electroweakinos up to 320 (700) GeV with $\Delta\sim 30$ GeV and up to 220 (620) GeV with $\Delta\sim 20$ GeV if gauginos decay via gauge bosons (light sleptons). We emphasize that the $3\ell$ channel is technically challenging for mass gaps below $\sim 12$ GeV.}

\begin{document} 
\maketitle
\flushbottom

%%%%%%%%%%%%%%%%%%
%%%%%%%%%%%%%%%%%%
\section{Introduction}

Low-energy supersymmetry (SUSY) is one of the main candidates for new physics (NP) beyond the Standard Model (SM). It is an important focus of the LHC new physics searches, which so far have only obtained null results in the 7+8 TeV run.

Light colored superparticles -- squarks and gluinos -- were thought to provide the best potential for the discovery. However, in the first three years of LHC run, the LHC searches
have excluded first two generation squarks and gluinos with masses below about 1 TeV~\cite{ATLAS:2013sma,ATLAS:2013062,ATLAS:2013tma,Chatrchyan:2013lya,CMS:2012yua}.
On the other hand, electroweak particles -- binos, winos, higgsinos -- might be much lighter than squarks and gluinos. 
Indeed, there are theoretical models predicting such a spectrum; models of split SUSY~\cite{ArkaniHamed:2004fb,Giudice:2004tc} and anomaly mediation~\cite{Randall:1998uk,Giudice:1998xp} are the most popular examples. Therefore, bino, wino and higgsino, collectively referred to as the electroweakinos,  have become the obvious target for LHC searches.

LHC is setting bounds on the SUSY electroweak sector indirectly through the cascade decays of squarks and gluinos~\cite{ATLAS:2013062,ATLAS:2013tma,Giudice:2010wb}. However, these bounds are quite model dependent and rely on having squarks and gluinos below the TeV scale. Additionally, several direct searches for electroweakinos have been performed, for which the electroweak particles are produced directly from Drell-Yan processes~\cite{ATLAS:2013rla,Chatrchyan:2013006,ATLAS:2013zka}.
The sensitivities of such direct searches depend on the mixtures and masses of the electroweakinos as well as on the possible existence of light sleptons. These searches can be generally classified depending on the nature of the Lightest Supersymmetric Particle (LSP).
If the LSP is not a gravitino, typically the best sensitivity is expected from multi-lepton channels coming from the decay of pair produced charginos/neutralinos~\cite{ATLAS:2013rla,Chatrchyan:2013006}. 
Here, the typical signatures are trilepton, same-sign dilepton, four lepton and sizable missing energy ($\MET$). 
If the LSP is a gravitino, additional signatures involving photons can be viable~\cite{Aad:2012jva,Chatrchyan:2012bba,Meade:2009qv}. 

Among the current multilepton searches, 
the best sensitivity is generally achieved in the $3\ell$ channel
~\cite{ATLAS:2013rla,Chatrchyan:2013006}, although improvements can be obtained from other multilepton modes for certain cases~\cite{Chatrchyan:2013006,ATLAS:2013zka}. In the $3\ell$ channel, the LHC sensitivity~\cite{ATLAS:2013rla,Chatrchyan:2013006} reaches up to about 320 (720) GeV if NLSP gauginos decay via intermediate gauge bosons (light sleptons) and the LSP is massless. However, going to more squeezed scenarios, 
the final state leptons are softer and the missing energy is smaller. Therefore, it is more difficult to successfully isolate the signal from the Standard Model backgrounds.  For example, for 30 GeV mass-gap between the chargino/second lightest neutralino and the LSP, only gauginos up to 115 (250) GeV are excluded in this channel.

One known way to improve the search sensitivity to small mass-gap regions is to utilize a hard initial state radiation (ISR)~\cite{Gunion:2001fu,Carena:2008mj,Drees:2012dd,Alvarez:2012wf,Alves:2012ft,Bartels:2012ex,Dreiner:2012gx,Delgado:2012eu,Berggren:2013vfa}. ISR photons were already used to search for nearly degenerate gauginos at LEP~\cite{Abreu:2000as,Heister:2002mn,Abbiendi:2002vz}.
At the LHC, the existence of a hard ISR jet and the resulting boost of the NLSP gauginos usually lead to the monojet plus missing energy topology. This signature is useful  when the mass-gap is very small.  However, the sensitivity of the monojet plus missing energy search drops quickly when the mass splitting becomes larger. 
 
In this work, we investigate the role of a ISR jet in the search for gauginos with moderate mass splitting $m(\chi_{{\rm{NLSP}}})-m(\chi_{{\rm{LSP}}})\sim (10-50)$ GeV.
In this region, the final state visible particles can be hard enough to be measured. They usefully encode information of the small-gap decay kinematics. 
We will specifically work in the $3\ell$ + $\MET$ channel arising from NLSP gaugino pair production.

The rest of this paper is organized as follows: In Sec.~\ref{sec:SignalBackground}, we present the details of our simulation of the signal and background processes. In Sec.~\ref{sec:physics}, we discuss the kinematic features that best distinguish the signal of the small-gap scenario from the SM background. In particular we show the relevance of the minimum value of the invariant mass of same-flavor opposite-sign leptons, the angular separation $\Delta \phi(j_1,\MET)$ between the missing energy and the ISR jet $p_T$, and  the ratio variables  $\frac{\MET}{p_T(j_1)}$ and $\frac{p_T(\ell_1)}{p_T(j_1)}$. In Sec.~\ref{sec:results} we present the optimization of our analysis and the exclusion bounds we get on the chargino parameter space, both at the 8 TeV LHC with 21/fb data and at the 14 TeV LHC with 300/fb. We reserve Sec.~\ref{sec:conclusions} for our conclusions. Finally, for completeness, we present two Appendices. Appendix~\ref{sec:app-mc} contains some more details on our event generation, and Appendix~\ref{sec:app-slep} compares the kinematic variables in the case of heavy sleptons and light sleptons.

%%%%%%%%%%%%%%%%%%
%%%%%%%%%%%%%%%%%%

%%
\section{Signal and background processes}\label{sec:SignalBackground}

We study the 3 lepton plus $\MET$ signature coming from the associated production of a chargino and a second lightest neutralino, 
followed by the decays $\chi_1^\pm\to\chi_1^0\ell\nu$ and $\chi_2^0\to \chi_1^0\ell\ell$. To have a easier comparison with the LHC searches, we follow closely the ATLAS assumptions in~\cite{ATLAS:2013rla}\footnote{Note that the CMS assumptions and analysis in~\cite{Chatrchyan:2013006} are similar.}. Namely, we assume that the LSP is bino-like and the NLSP gauginos are mainly wino and degenerate in mass. 

We consider two separate scenarios depending on how gauginos decay to leptons. In the first scenario, gauginos decay via off-shell gauge bosons as
\beq
\chi^\pm_1 \to W^{(*)} \chi^0_1 \to \ell \nu \chi^0_1, \quad \chi^0_2 \to Z^{(*)} \chi^0_1 \to \ell \ell \chi^0_1, \qquad (\ell \= e,\mu).
\eeq
In this case, the relevant leptonic branching fraction of $\chi^\pm_1 \chi^0_2$ is small about $1.5\%$, since it is suppressed by  the gauge boson leptonic branching fractions. 

In the second scenario, we assume that the left-handed sleptons and sneutrinos are light with masses given by $m_{\tilde\ell_L}=m_{\tilde\nu}=(m_{\chi^\pm_1}+m_{\chi^0_1})/2$. Therefore, they give the main contribution to the gaugino decays to first two generation sleptons
\beq
\chi^\pm_1 \to \tilde{\ell} \nu, \, \tilde{\nu} \ell \to \ell \nu \chi^0_1, \quad \chi^0_2 \to \tilde{\ell} \ell \to \ell \ell \chi^0_1, \qquad (\ell \= e,\mu).
\eeq
 The relevant leptonic branching ratio is much larger,  BR(leptonic)$\sim 0.3$.

The main difference between the two scenarios is given by the leptonic branching ratios. As discussed in detail in the Appendix \ref{sec:app-slep}, the distributions for the kinematic variables that we will utilize in this paper are quite similar in these two scenarios. For this reason, we will simply use the first scenario to discuss the physics throughout this paper. However, as leptonic branching ratios differ, in Sec.~\ref{sec:results}, we will present a separate  exclusion reach for the case of light sleptons.

The dominant irreducible SM background to the $3\ell$ + $\MET$ final state is the $WZ$ diboson production~\cite{ATLAS:2013rla,Chatrchyan:2013006}\footnote{By $WZ$, we refer to all $3\ell$ production processes mediated by on/off-shell $Z^{(*)}$ as well as by $\gamma^{(*)}$.}. Additional backgrounds include other diboson productions of $ZZ$ and $WW$, single- and pair-production of top quarks, and single $W$ or $Z$ boson production in association with jets or photons. In our work, we consider only the $WZ$ process. The $WZ$ process has a decay topology similar to that of the signal as $(\ell, \nu)$ come from one side of the decay chain while $(\ell^+, \ell^-)$ come from the other side.
The ATLAS analysis~\cite{ATLAS:2013rla} shows in fact that other irreducible backgrounds generally amount to at most $\sim 30\%$ of the $WZ$ background in all the relevant signal regions. Along our discussion, we comment on how the $ZZ$ background among others can be suppressed in our work.

We model signal and background using MadGraph5~\cite{Alwall:2011uj} interfaced with
Pythia 6.4~\cite{Sjostrand:2006za} for parton showering. We allow up to one additional parton in the final state, and adopt the MLM matching scheme~\cite{Mangano:2006rw}. We cluster jets using the anti-$k_T$ algorithm~\cite{Cacciari:2008gp} implemented in FastJet-2.4.3~\cite{Cacciari:2011ma} with a radius parameter $R=0.4$. We compute the NLO cross sections for the signal and the background processes using {\sc Prospino}~\cite{Beenakker:1996ed} and {\sc MCFM}~\cite{Campbell:2010ff}, respectively. Furthermore, we normalize our background in such a way to match the background presented in the ATLAS paper~\cite{ATLAS:2013rla}. In Appendix~\ref{sec:app-mc}, we further validate our background event sample.

We do not carry out a specific detector simulation. However, it is important to take into account the lepton identification efficiency and the lepton-photon isolation. In this work,  we use $p_T$-dependent lepton identification efficiencies taken from Refs.~\cite{Aad:2011mk,Aad:2009wy}. In particular, we note that the electron efficiency drops  below $70\%$ for $p_T^e \lesssim {\cal O}(10)$GeV and instead the muon identification efficiency remains high ($\sim 90\%$ for $p_T^\mu \gtrsim 7$ GeV).
In Appendix \ref{sec:app-mc}, we collect the details of the lepton identification efficiency and discuss the importance of lepton and photon isolation.

%%%%%%%%%%%%%%%%%%%%%%%%%%%%%%%%%%%%%%%%%%%%%%%%%%%%%%%%%%%%%%%%%%%%%%%%%%%%%%%%%%%%%%%%%%%%%%%%%%%%%%%%%
\section{Kinematics of small-gap gaugino decays} \label{sec:physics}

The scenarios we are interested in are characterized by a small mass difference between the mother and one of the daughter particles, with the other SM daughter particle approximately massless. 
This configuration is realized when the NLSP gauginos are close in mass to the LSP:  $\Delta \equiv m(\chi_{{\rm{NLSP}}})-m(\chi_{{\rm{LSP}}}) \ll m(\chi_{{\rm{NLSP}}})$.
In this section, we discuss the various kinematic features of this small-gap region that can be used to distinguish the signal from the $WZ$ background. 

\subsection{Basic kinematic variables}

\begin{figure}[t] \centering
\includegraphics[width=\textwidth]{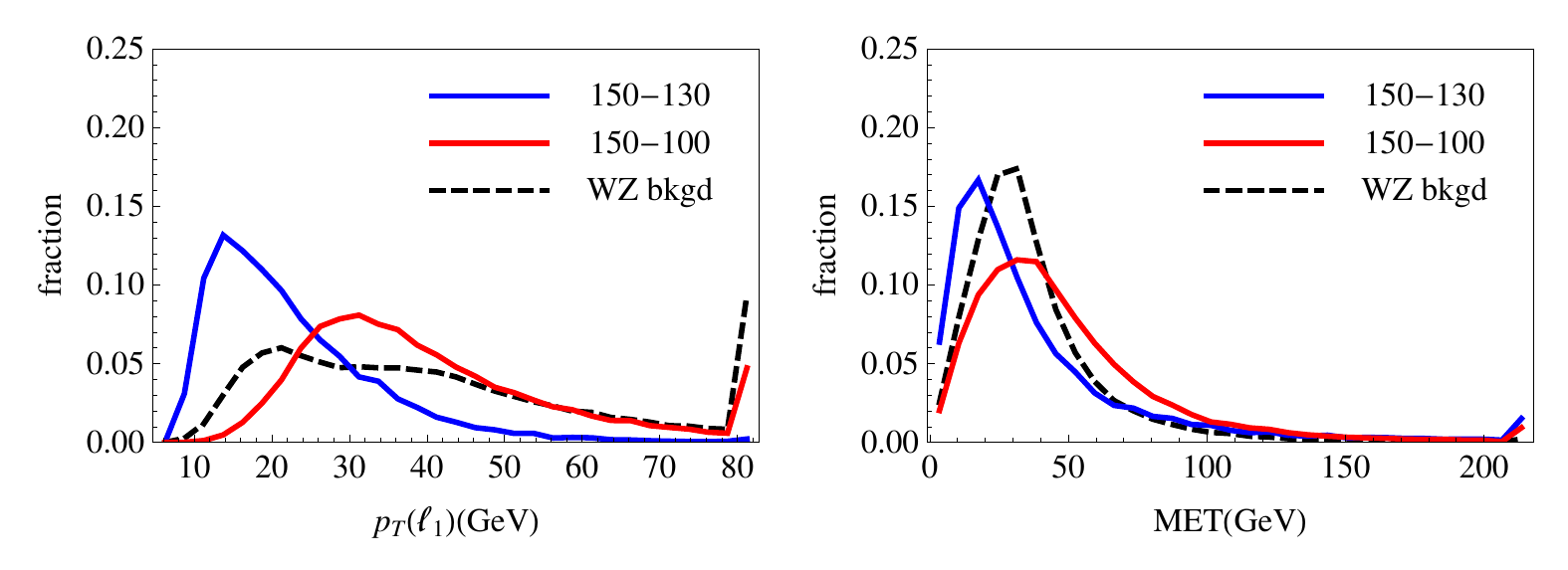}
\caption{
Hardest lepton $p_T$ (left) and $\MET$ (right) spectrum at the 8 TeV LHC. The blue (red) curve represents the scenario with $m_{\chi^0_2}=m_{\chi^\pm_1}=150$ GeV and $m_{\chi^0_1}=130\, (100)$ GeV. The black dashed line represents the background distribution. Curves are normalized to one. The last bin contains all the events outside of the plot range. For illustrative purposes, we only show events satisfying the Loose-$p_T$ baseline cuts with $p_T(\ell)>7$ GeV and $p_T(j_1)>30$ GeV  presented in Table~\ref{tab:150130}.
}
\label{fig:discrim-ptl1}
\end{figure}

It is well-known that leptons coming from the decay in the small-gap scenario are soft. The smaller the gap, the softer are the leptons. \Fig{fig:discrim-ptl1} shows this dependence on the mass-gap for two benchmark scenarios with different mass-gaps: $m_{\chi^0_2}=m_{\chi^\pm_1}=150$ GeV and $m_{\chi^0_1}=130\,(100)$ GeV in blue (red). Soft leptons have been used as a discriminator not only in theoretical studies~\cite{Rolbiecki:2012gn,Delgado:2012eu,Giudice:2010wb} but also  in LHC analysis~\cite{ATLAS:2013rla}. However, for $\Delta \sim (30-50)$ GeV which is also part of the parameter space we are interested in, the signal lepton spectrum becomes similar to the background lepton spectrum; see the distribution in black on the left panel of \Fig{fig:discrim-ptl1}. 
Unlike leptons, LSPs from gaugino decays carry most of the gaugino's energy-momentum as they are much heavier. Nonetheless, events arising from small gap scenarios do not lead to a large missing energy ($\MET$). The right panel of \Fig{fig:discrim-ptl1} shows that the $\MET$ from two ${\cal O}(100\,{\rm{GeV}})$ missing particles is not larger than the SM $\MET$. This is  because the two LSPs are usually produced in back-to-back configuration.

The invariant mass of the same-flavor opposite-sign (SFOS) lepton pair, denoted by $\mSFOS$, is another useful variable.  The invariant mass of SFOS leptons from $Z$ boson decays has a distribution peaked at $m_Z$ while, for small-gap scenarios, the SFOS pair from neutralino decays is kinematically limited to have a small invariant mass forming an edge at around $\Delta$. Since $m_Z \gg \Delta$, vetoing events with large $\mSFOS$ is useful to reject a large fraction of the background from on-shell $Z$ decays, while still keeping a large signal acceptance.

\begin{figure}[t] \centering
\includegraphics[width=\textwidth]{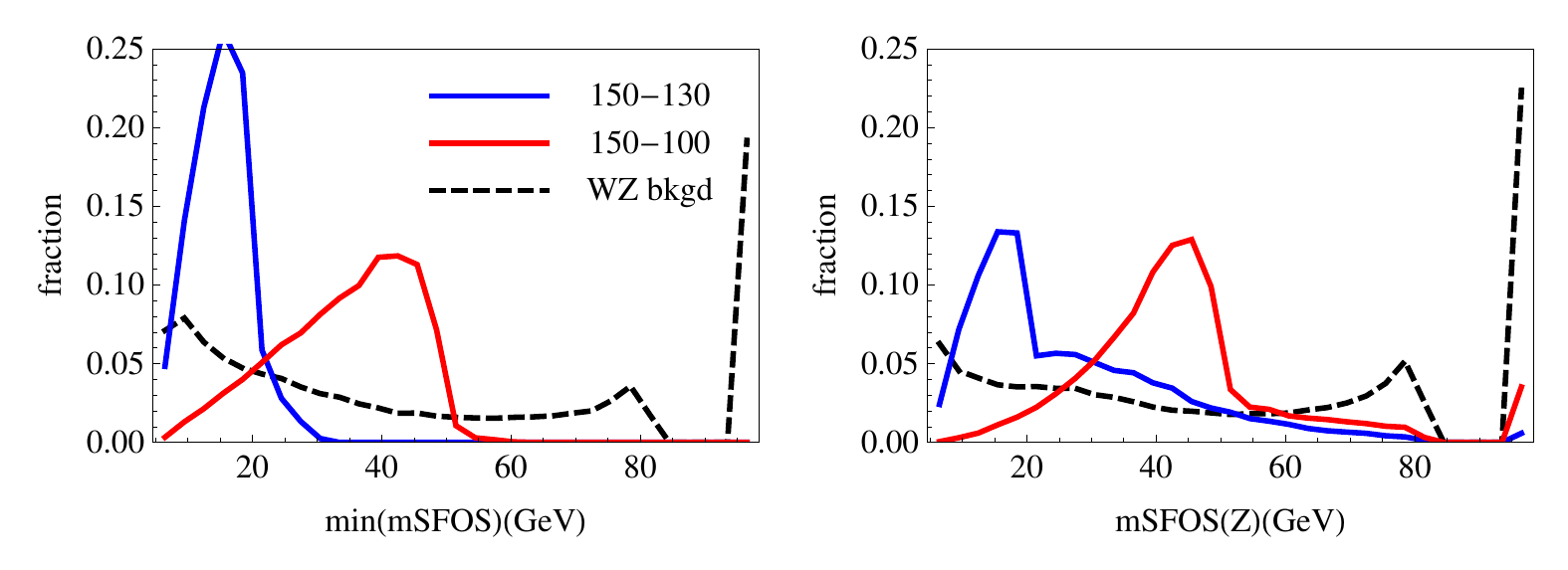}
\caption{Distribution for the two variables min($\mSFOS$) (left) and $\mSFOS(Z)$ (right) presented in the text. min($\mSFOS)$ has a clear edge at around $\Delta$. The baseline cut on min$(\mSFOS)$ is relaxed to min$(\mSFOS)>2$ GeV for illustrative purposes. The other details of the plots are as for \Fig{fig:discrim-ptl1}.} 
\label{fig:mSFOS}
\end{figure}

A complication arises when there are two possible SFOS lepton pairs. The ATLAS and CMS $3\ell$ gaugino searches~\cite{ATLAS:2013rla,Chatrchyan:2013006} use the $\mSFOS$ closest to $m_Z$, denoted by $\mSFOS(Z)$, to choose a correct pair from $Z$ boson decays. We find that an additional useful variable is the minimum of all possible invariant masses denoted by min($\mSFOS$)\footnote{min($\mSFOS)$ has been used in a different context in Ref.~\cite{Rolbiecki:2012gn}.}. The min$(\mSFOS)$ has a clearer edge at $\Delta$ than $\mSFOS(Z)$ (see \Fig{fig:mSFOS} for comparison). This is because, for signal events, the correct pair invariant mass is always smaller than $\Delta\, (\ll m_Z)$, whereas uncorrelated wrong pairs tend to have a larger invariant mass closer to $m_Z$. In Sec.~\ref{sec:results}, we will use a upper cut on min$(\mSFOS)$ to better utilize the edge of small-gap signals.

We note, however, that focusing on events with small SFOS invariant mass is not always advantageous. As shown in \Fig{fig:mSFOS}, the background also has an accumulation of events at low invariant masses, mainly coming from the virtual photon process. 
On the other hand, the signal distribution for min$(\mSFOS)$ 
 peaks close to the edge at $\sim \Delta$. 
Therefore, a lower cut on min$(\mSFOS)$ can also help rejecting the off-shell photon background. Furthermore, excluding very small values for min($\mSFOS$) efficiently suppresses the background coming from low mass hadronic resonances decaying to leptons~\cite{ATLAS:2013rla,Chatrchyan:2013006}. 

\subsection{Useful variables in the presence of the ISR jet}

\begin{figure}[t] \centering
\includegraphics[width=0.48\textwidth]{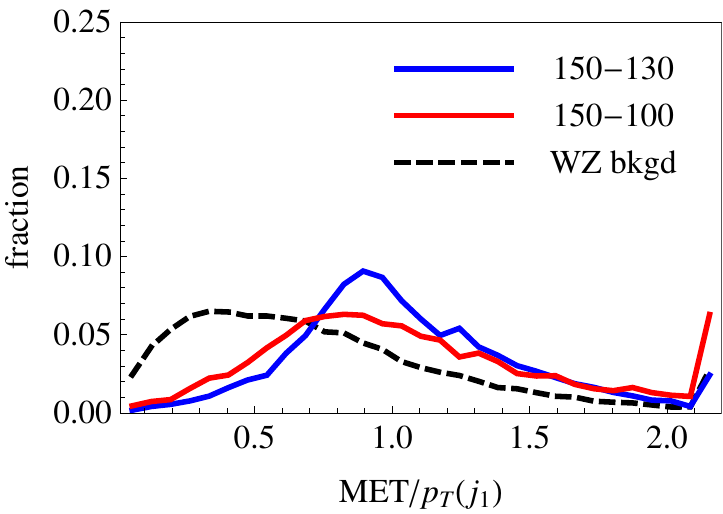}
\includegraphics[width=0.48\textwidth]{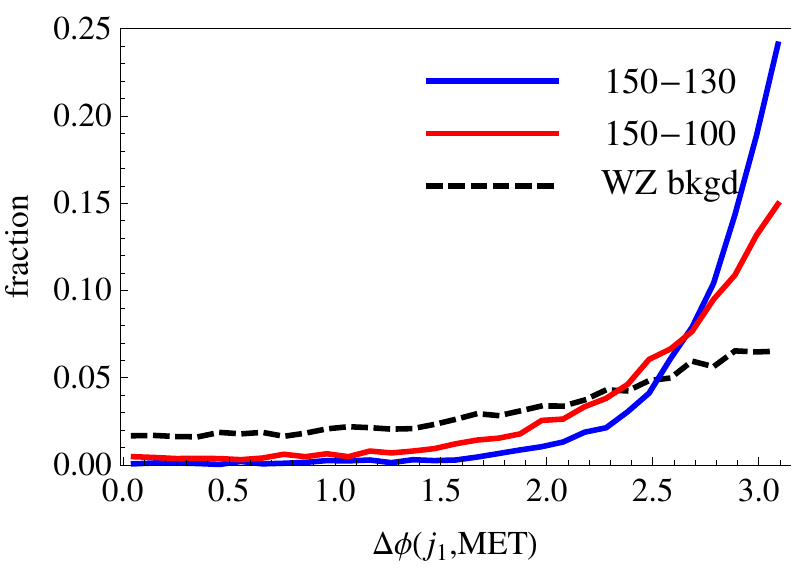}
\caption{Right: $\MET/p_T(j_1)$ spectrum. Left: $\Delta\phi(j_1,\MET)$ spectrum. The other details of the plots are as for \Fig{fig:discrim-ptl1}.}
\label{fig:discrim-ptl1ptj1}
\end{figure}

We now discuss the kinematics in the presence of an energetic ISR, $j_1$. 
The most significant effect is on the missing energy, which can be estimated as 
\beq\label{eq:METboost}
-\vec{E}_T^{\rm miss} = \vec{p}_T(j_1) + \sum \vec{p}_T(\ell), \qquad  |\vec{p}_T(\ell)| \sim \gamma E_{\ell}^0.
\eeq
For the signal, $E_{\ell}^0 \sim \Delta$ and is small in the limit we are considering. Therefore, ${E}_T^{\rm miss}$ is strongly correlated with $p_T(j_1)$. On the other hand, for the background, the contribution from the second term of \Eq{eq:METboost} is also important, and thus the correlation is weaker.  
To effectively encode this feature, we introduce a ratio variable
$\MET/p_T(j_1)$
in our analysis. The distributions for this variable for two signal scenarios and for the background are shown in the left panel of \Fig{fig:discrim-ptl1ptj1}. As expected from \Eq{eq:METboost}, $\MET/p_T(j_1)$ peaks at around 1 in the case of the signal. This feature is independent on the overall SUSY mass scale and holds generically in the case of small gap scenarios; thus, it can be used to search for a wide range of parameter space. We also observe that, due to a partial cancellation between the $p_T$ of the ISR and those of the leptons, the background distribution peaks at values smaller than one\footnote{We note that requiring a correlation between $\MET$ and the ISR $p_T$ significantly suppresses $ZZ$ backgrounds because the dominant source of $\MET$ is mis-measurement of momentum there.}.

Following the same argument, we can also conclude that, for the signal, the direction of the missing momentum is strongly correlated with (opposite to) the direction of the ISR jet. The smaller the mass gap, the stronger the correlation, as the second term in \Eq{eq:METboost} becomes less important. On the other hand, the correlation should be much weaker in the background.  All these features are clearly visible in the right panel of \Fig{fig:discrim-ptl1ptj1}. Therefore, $\Delta\phi(j_1,\MET)$\footnote{Ref.~\cite{Alvarez:2012wf} has considered a similar variable in a different context.} defined as angular separation in the transverse plane is another useful variable we can use to discriminate between signal and background.

\begin{figure}[t] \centering
\includegraphics[width=0.48\textwidth]{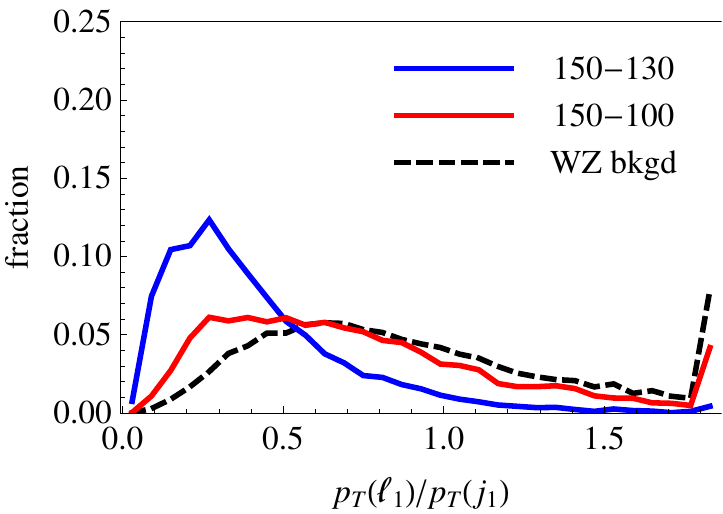}
\caption{$p_T(\ell_1)/p_T(j_1)$ spectrum. The other details of the plots are as for \Fig{fig:discrim-ptl1}.}
\label{fig:discrim-dphij1met}
\end{figure}

Next, we consider the effect of the ISR on the kinematics of the leptons. As a rough estimate, we have
\beq
p_T ( \ell) \sim \gamma E_\ell^0, \qquad \gamma \sim \frac{\sqrt{p^2_T (j_1)/4 + M^2} }{M}.
\label{eq:boost}
\eeq
For the signal, $M = m(\chi_{\rm NLSP})$ and  $E_\ell^0 \sim \Delta $. For the $WZ$ background, we have $M = m_{\rm W/Z}$ and $E_\ell^0 = m_{\rm W/Z} /2$. Since $m(\chi_{\rm NLSP}) > m_{\rm W/Z}$ for the parameter space of interest, we expect that, for a given $p_T (j_1)$, the leptons from the background will get a larger boost.

This observation leads us to consider a ratio variable between the ISR jet $p_T$ and the lepton $p_T$: 
$\frac{p_T(\ell_1)}{p_T(j_1)}$ (see \Fig{fig:discrim-dphij1met} for the distribution). We note that, in order not to loose much of the signal acceptance, the ISR we will consider in this work is not very hard ($p_T(j_1)\gtrsim 30$ GeV). Therefore, the boost is not a large effect on this variable, and the performance of this variable alone is similar to that of a hard lepton veto. However, when combined with the other variables we consider in this work, it performs better. It is worthwhile to notice that the use of this variable is especially powerful at the LHC high luminosity stage, at which, not being limited by statistics, we can devise a harder cut on these variables. We will discuss this feature at the 14 TeV LHC with 300/fb data in the next section.

%%%%%%%%%%%%%%%%%%
%%%%%%%%%%%%%%%%%%
\section{LHC reach} \label{sec:results}

\subsection{Optimization for $\Delta \sim (30-50)$GeV at LHC8 21/fb}\label{sec:res-3050}

Based on the features of the small-gap region discussed in the previous section, we use the following five variables to optimize our analysis
\beq
\MET, \quad p_T(\ell), \quad \frac{p_T(\ell_1)}{p_T(j_1)}, \quad \frac{\MET}{p_T(j_1)}, \quad \Delta\phi(j_1,\MET).
\label{eq:var-opt} \eeq
A common lepton $p_T$ cut ($p_T(\ell)$) is applied to all three leptons although more dedicated cuts for each leptons could be derived. 
We do not include min$(\mSFOS)$ in the list because the lower and upper cuts on this variable are in large part determined by the mass-gap of the benchmark considered.

We apply baseline selection cuts that are chosen to resemble the current ATLAS baseline cuts~\cite{ATLAS:2013rla}:
\bei
\item exactly $3\ell$: $p_T>10$ GeV, $\eta < 2.5$,  min$(\mSFOS)>12$ GeV, and $\mSFOS(Z)<81$ GeV.
\eei
Most events passing these lepton cuts also pass the lepton trigger requirements used by ATLAS~\cite{ATLAS:2013rla}. Furthermore, we ask for a jet:
\bei
\item ISR jet with $p_T>30$ GeV, $\eta < 2.5$. 
\eei
In addition, we impose
\bei
\item 18 GeV$<$min$(\mSFOS)<\Delta$. This cut is varied depending on the mass-gap $\Delta$.
\eei
Note that we impose min$(\mSFOS)>18$ GeV, which is  harder than the ATLAS's  requirement of 12 GeV. The 12 GeV lower bound is applied to avoid the background
from low-mass hadronic resonances. As discussed in the previous section, some lower cut can also be useful to additionally suppress the virtual photon background. We find that 18 GeV is the optimal cut for mass gaps in the interval
$\Delta \sim (30-50)$ GeV.

The optimization is carried out to maximize the conveniently chosen significance estimator
\beq
\sigma \= \frac{S}{\sqrt{B + (0.15B)^2}}
\label{eq:sigest} \eeq
where $S$ and $B$ are the number of signal and background events at a given luminosity. In the denominator, the simple estimate of the statistical uncertainty, $\sqrt{B}$, and the approximate systematic uncertainty attributed to 15\% of the background are added in quadrature. Furthermore, in the process of optimizing, we require at least 10 signal events at a given luminosity to ensure minimal statistics.

In the following, we present the optimization for the scenario $m_{\chi_1^\pm}=m_{\chi_2^0}=150$ GeV, $m_{\chi_1^0}=120$ GeV, denoted by ``(150-120)'', at the 8 TeV LHC with 21/fb data. We name these optimal cuts ``Tight-$p_T$ cuts'':
\beq
\qquad \MET>20\GeV,\, p_T(\ell)<50\GeV,\, \frac{p_T(\ell_1)}{p_T(j_1)} < 1.21,\, \frac{\MET}{p_T(j_1)}>0.64,\, \Delta\phi(j_1,\MET)>2.4. 
\eeq
The cut-flow for these optimal cuts is shown in Table.~\ref{tab:150120}. In Table.~\ref{tab:150120}, we also show the ATLAS results that we obtain by simulating the ATLAS signal region named ``SRnoZa'' in Ref.~\cite{ATLAS:2013rla}. The ATLAS SRnoZa selection is in fact the one that is most sensitive to small gap scenarios. As we can observe from the table, our cuts can improve the ATLAS results. Interestingly enough, even the lower and the upper cuts on min$(\mSFOS)$ alone can already improve the ATLAS results. The additional variables associated to the ISR jet can give a larger enhancement of the signal significance.

\begin{table}[t] \centering
\begin{tabular}{c|l|c|c|c|c}
\hline \hline
(150-120)  & cuts & $S $ &  $\frac{S}{B}$ &  $\frac{S}{\sqrt{B}} $  &  $\frac{S}{\sqrt{B + (0.15\cdot B)^2}}$  \\
\hline \hline
 & $p_T(\ell)>$10 GeV, $p_T(j)>$30 GeV, & 18 & 0.17 & 1.8  & 0.97 \\
Tight-$p_T$ baseline& min$(\mSFOS)>18$ GeV,  & & & & \\
& $\mSFOS(Z)<$81 GeV  & & & & \\
\hline
& min$(\mSFOS)<\Delta$  & 17 & 0.47 & 2.8 & 2.1\\
Tight-$p_T$& $\Delta \phi(j_1,\MET) >$ 2.4 & 14 & 0.91 & 3.5 & 3.1 \\
cuts & $\MET/p_T(j_1)>$0.64 & 12  & 1.4 & 4.1 & 3.7 \\
& $ \bmat \MET>20\GeV,\, p_T(\ell_1)<50 \GeV \\\hspace{-0.5cm} p_T(\ell_1)/p_T(j_1)<1.21 \emat  $  & 11 & 1.7 & 4.3 & 4.0 \\
\hline \hline
ATLAS search~\cite{ATLAS:2013rla}& SRnoZa  & 17 & 0.32 & 2.3 & 1.6 \\
\hline \hline
\end{tabular}
\caption{Cut-flow with Tight-$p_T$ cuts for the $(150-120)$ benchmark at the 8 TeV LHC with 21/fb data. The ATLAS results from the SRnoZa selection, the signal region most sensitive to the small-gap region, is also shown for comparison. In particular, SRnoZa requires $\mSFOS(Z)<60$ GeV,  min$(\mSFOS)>12$ GeV, $\MET>50$ GeV and either $\MET<75$ GeV or $m_T(W)<110$ GeV 
or $p_T(\ell_3)<30$ GeV.}
\label{tab:150120}\end{table}

\begin{figure}[t] 
\centering
\includegraphics[width=\textwidth]{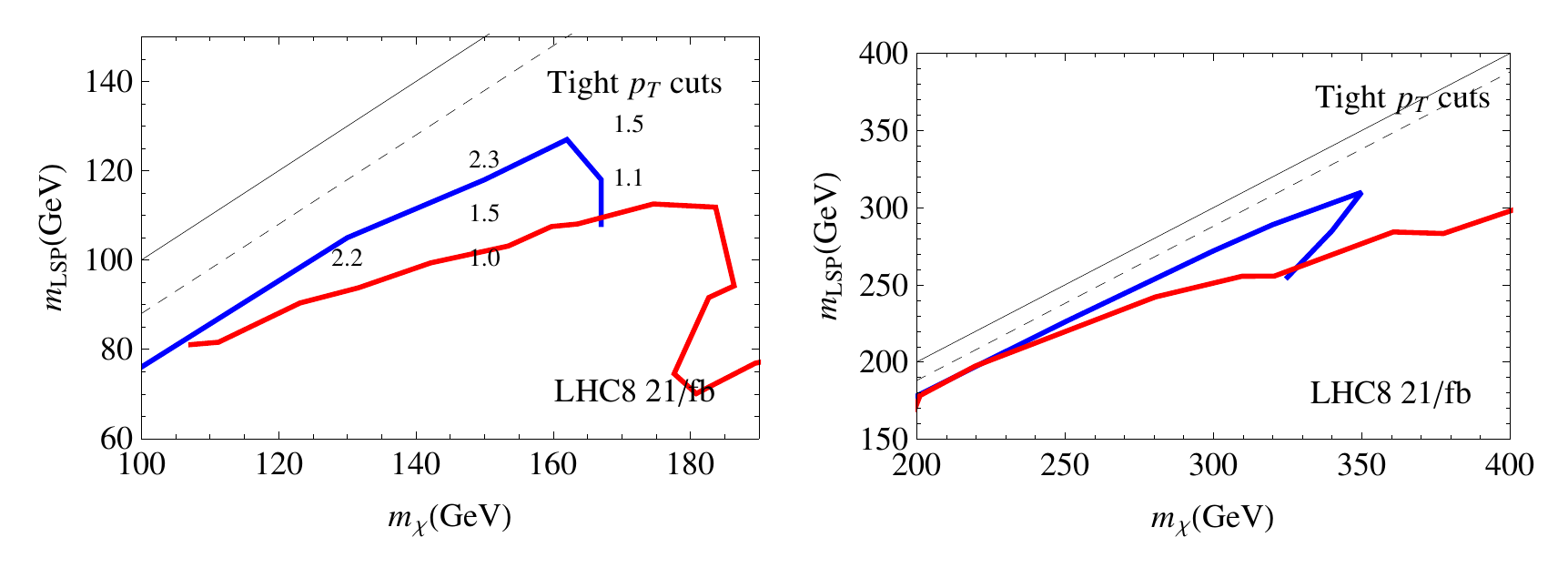}
\caption{Expected exclusion reach with ``Tight-$p_T$ cuts'' at LHC8 21/fb. The two scenarios shown are gaugino decays via gauge bosons (left) or via on-shell light sleptons (right). We present the expected reach using our cuts in blue, and the ATLAS official boundary in red for comparison. In the left panel, we also show the ratio of our significance to that of ATLAS SRnoZa. The solid gray line is for $m_{\rm{LSP}}=m_{\rm{NLSP}}$, and the dashed gray line is where $\Delta =12$ GeV. }
\label{fig:reach8-tight}
\end{figure}

Although the optimal cuts for a benchmark scenario are not optimal in the entire parameter space under consideration, 
we use the ones obtained for the $(150-120)$ benchmark to study the LHC reach with 21/fb 8 TeV data.
The expected exclusion reach using the Tight-$p_T$ cuts is presented in \Fig{fig:reach8-tight}, both for the model with heavy sleptons (left panel) and the one with light sleptons mediating the leptonic decays of the gauginos (right panel). As discussed in more detail below, a useful measure of the power of our method is the ratio of the signal significance obtained by using our cuts and the one by using ATLAS cuts. These are shown in the left panel of  \Fig{fig:reach8-tight} for some representative points in the parameter space. 

Before closing this section, we elaborate on how we arrived to the estimation of the reach. We have made several simplifications in our simulation. As explained above, we have only simulated the $WZ$ background, while ignoring the other irreducible and reducible backgrounds. Furthermore, we did not carry out a detailed detector simulation. To partially take into account these additional effects, we follow the following procedure.    To determine which region of parameter space can be probed at a given luminosity, we compare the signal significance obtained using our cuts to the one we obtain by simulating the ATLAS analysis for points at the boundary of its exclusion region. We find that a constant significance of $\sigma \simeq 3.5$  approximately reproduces ATLAS's official boundary in small-gap region both in the case of heavy and light sleptons. Therefore, we estimate that scenarios with a signal significance about $\sigma\simeq 3.5$ under our cuts will be approximately the limit we can expect to probe. For example, 
following this criterion, we conclude that the (150-120) benchmark, which has $\sigma \sim 4$, is close to the border of our expected reach at the 8 TeV LHC with 21/fb data. This is how the exclusion boundary in \Fig{fig:reach8-tight} is obtained.  However, we emphasize that this procedure is just an estimate of the effects that we have not taken into account, as the efficiencies and systematics can clearly depend on the particle spectrum.  A more robust measure of the improvement we achieve is the ratio of the signal significance obtained by using our cuts and that obtained using the ATLAS SRnoZa cuts. This ratio is also presented in the left panel of \Fig{fig:reach8-tight}.

\subsection{Challenges for smaller mass gaps $\Delta\lesssim 30$ GeV}

The sensitivity of our analysis drops rapidly at smaller mass gaps $\Delta \lesssim 30$ GeV. For example, for the scenario $m_{\chi_1^\pm}=m_{\chi_2^0}=150$ GeV, $m_{\chi_1^0}=130$ GeV (denoted by ``(150-130)'') with $\Delta = 20$ GeV, the efficiency under baseline cuts is already only about $0.8\%$ corresponding to 6 events at 21/fb. The biggest drop in efficiency  is due to the requirement 
of min$(\mSFOS)>12$ GeV and the cut on the $p_T$ of the leptons. On the other hand, small invariant masses and soft leptons were the characteristic features of the small-gap region which successfully discriminated between signal and background. Can we adjust the corresponding cuts to more squeezed regions? 

It might be technically difficult to relax the min$(\mSFOS)>12$ GeV cut since this cut is needed to avoid low-mass hadronic bound states~\cite{ATLAS:2013rla,Chatrchyan:2013006}. 
Therefore, to lower down the min$(\mSFOS)$ cut one should take 
into account this additional source of background.  We do not attempt to follow this route. 
One possibility to avoid using the min$(\mSFOS)$ variable would be to utilize a different final state not having SFOS leptons. One such example is the same-sign dilepton channel. Although this channel maybe useful in certain scenarios, we focus on the three-lepton final state in this paper. 

Being able to use softer leptons can also enhance the sensitivity. But, lower thresholds of lepton momentum are usually avoided because lepton identification becomes inefficient while the jet fake rate increases. Electron efficiency especially drops below 70\% for $p_T \lesssim {\cal O}(10)$GeV~\cite{Aad:2011mk} (see more in Appendix~\ref{sec:app-mc}). Furthermore, leptons are currently used in the trigger of the analysis, too. The ATLAS $3\ell$ analysis~\cite{ATLAS:2013rla}, for example, uses lepton triggers based either on a single hard lepton of $p_T \ge 25$ GeV or soft dileptons with $p_T\ge 14\GeV$, or $p_T \ge 10\GeV-18\GeV$ in the case of muons and $p_T \ge 10\GeV-25\GeV$ in the case of electrons. Thus, it is not clear whether the analysis can be easily adapted to use softer leptons.

\subsection{Prospects at LHC8 and LHC14 with relaxed lepton $p_T$ thresholds} \label{sec:res-20}

Despite of these difficulties, it might still be feasible to use softer leptons in the analysis. Ref.~\cite{Sfyrla:2013yva,Zarzhitsky:2008zz}, for example, examined whether softer $p_T(\ell)$ thresholds can still suppress fake backgrounds and can be used in the $3\ell$ channel. Leptons as soft as (4-7) GeV are suggested in these analysis.
Furthermore, the lepton trigger may also be replaced or supplemented by the relatively hard jet that is required in our analysis. 
We also note that muon identification efficiency is still high $(\gtrsim 90\%)$ for muons as soft as 7 GeV with a low fake rate of ${\cal O}(1)$\% ~\cite{Aad:2009wy}. This feature may also suggest an analysis based on the requirement of at least two (soft) muons.

\begin{figure}[t] \centering
\includegraphics[width=\textwidth]{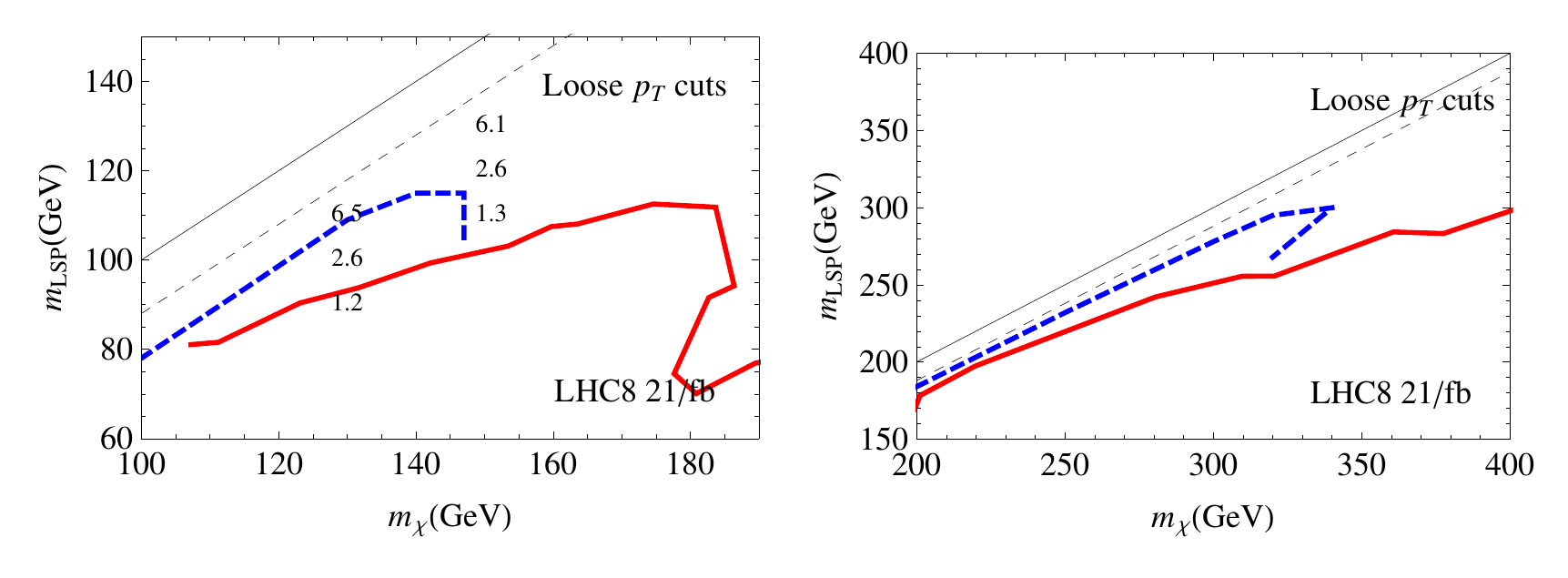}
\caption{Expected exclusion reach with ``Loose-$p_T$ cuts'' at LHC8 21/fb. The two scenarios shown are gaugino decays via gauge bosons (left) or on-shell sleptons (right).  The other details of the figure are as described in \Fig{fig:reach8-tight}.}
\label{fig:reach8-loose}
\end{figure}

In this subsection, we discuss the effects of using softer leptons at both the current LHC dataset and at the 14 TeV LHC. 
We first study the exclusion reach from the current dataset of 21/fb at LHC8, if the lepton threshold is relaxed to $p_T(\ell)>7$ GeV. 
In particular we optimize the variables in \Eq{eq:var-opt} for the (150-130) benchmark scenario. Since in this case the mass gap is very small, we also find it useful to relax the baseline cut to min$(\mSFOS)>12$ GeV (instead of 18 GeV). We find the following ``Loose-$p_T$ cuts'' to have the highest significance for this benchmark
\beq
\MET>20\GeV,\, p_T(\ell_1)<40\GeV,\, \frac{p_T(\ell_1)}{p_T(j_1)}<0.9,\,\frac{\MET}{p_T(j_1)}>0.64,\,\Delta\phi(j_1,\MET)>2.5.
\label{eq:cutset2}\eeq
These cuts are similar to the Tight-$p_T$ cuts found in Sec.~\ref{sec:res-3050}. The main difference is the lower lepton threshold, $p_T(\ell_1)<40$ GeV instead of $p_T(\ell_1)<50$ GeV. The corresponding cut flow is displayed in Table.~\ref{tab:150130}, and the expected reach with the current LHC dataset is shown in \Fig{fig:reach8-loose}. Comparing \Fig{fig:reach8-loose} to \Fig{fig:reach8-tight}, we observe that relaxing the lepton threshold allows to better probe the small-gap region close to $\Delta=12$ GeV. After the Loose-$p_T$ baseline cuts, in fact, the signal efficiency is $\sim 2.3\%$, to be compared with the $\sim 0.8\%$ we obtain with the Tight-$p_T$ baseline cuts. However, as already noted, we also observe that it is still difficult to probe 
the $\Delta \lesssim 12$GeV region due to the inevitable loss of signal efficiency from the cut min$(\mSFOS) >12$ GeV.

\begin{table}[t] \centering
\begin{tabular}{c|l|c|c|c|c}
\hline \hline
150-130  & cuts & $S $ &  $\frac{S}{B}$ &  $\frac{S}{\sqrt{B}} $  &  $\frac{S}{\sqrt{B + (0.15\cdot B)^2}}$  \\
\hline \hline
Loose-$p_T$baseline & $p_T(\ell)>$7 GeV, $p_T(j)>$30 GeV,  & 16 & 0.093 & 1.2  & 0.55 \\
& min$(\mSFOS)>12$ GeV, & & & & \\
&  $\mSFOS(Z)<$81 GeV & & & & \\
\hline
& min$(\mSFOS)<\Delta$  & 14 & 0.24 & 1.8 & 1.2 \\
Loose-$p_T$& $\Delta \phi(j_1,\MET)>$2.5 & 12& 0.52& 2.4 & 2.0\\
cuts & $\MET/p_T(j_1)>$0.64 & 11  & 0.72 & 2.8 & 2.4 \\
& $\bmat \MET>20\GeV,\, p_T(\ell_1)<40\GeV \\ p_T(\ell_1)/p_T(j_1)<0.9 \emat$  & 10 & 0.88 & 3.0 & 2.6 \\
\hline \hline
ATLAS search~\cite{ATLAS:2013rla} & SRnoZa  & 4.4 & 0.08 & 0.60 & 0.40 \\
\hline \hline
\end{tabular}
\caption{Cut-flow with Loose-$p_T$ cuts for the $(150-130)$ scenario at LHC8 21/fb. The other details of the table are as described in Table~\ref{tab:150120}.}
\label{tab:150130}\end{table}

\begin{figure}[t] \centering
\includegraphics[width=\textwidth]{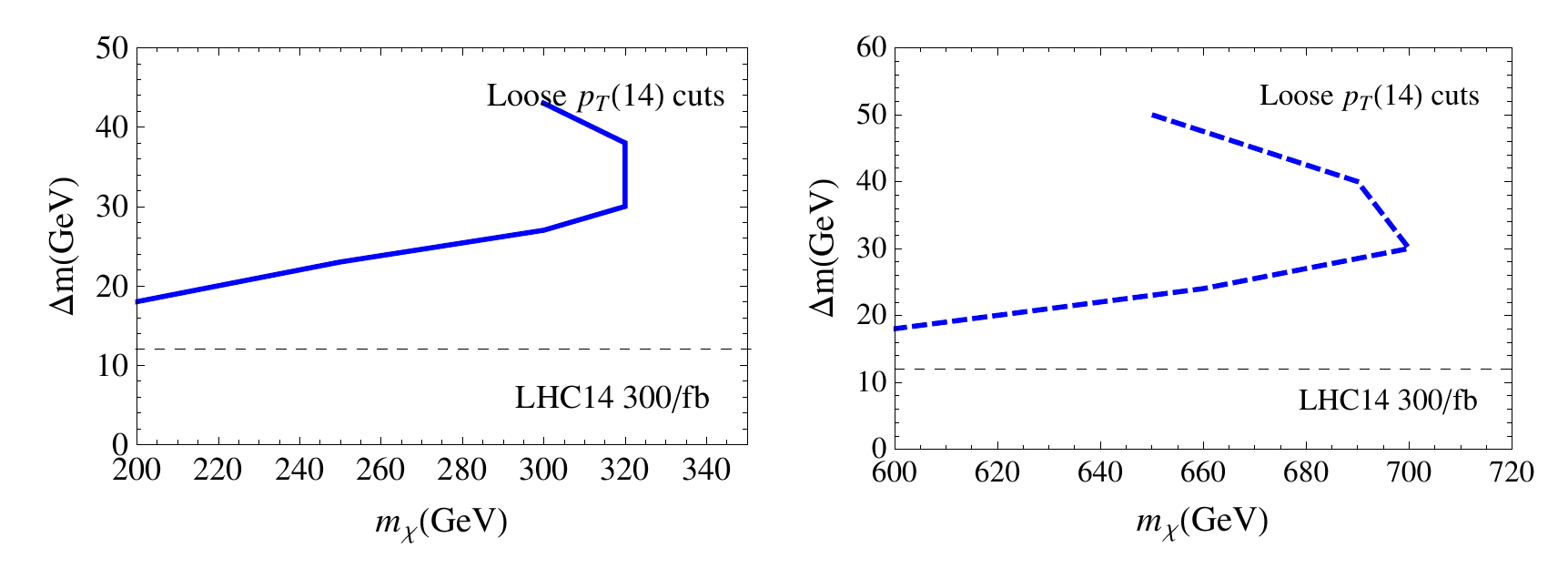}
\caption{Expected exclusion reach with ``Loose-$p_T$(14) cuts'' at LHC14 300/fb. The two scenarios shown are gaugino decays via gauge bosons (left) or on-shell sleptons (right).  The other details of the figure are as described in \Fig{fig:reach8-tight}.}
\label{fig:reach14}
\end{figure}

At the 14 TeV LHC with 300/fb data, signal and background rates are much higher, and the optimization of the cuts yields somewhat different results. Using $p_T(\ell)>7$ GeV in the baseline cuts, we obtain the ``Loose-$p_T$(14) cuts'' optimized for the (300-280) benchmark scenario ($m_{\chi_1^\pm}=m_{\chi_2^0}=300$ GeV and $m_{\chi_1^0}=280$):
\beq
\quad \MET>60\GeV, \quad p_T(\ell)<50\GeV, \quad \frac{p_T(\ell_1)}{p_T(j_1)} < 0.2, \quad \frac{\MET}{p_T(j_1)}>0.9. 
\eeq
We note that, for the 14 TeV LHC, we do not achieve a sizable improvement by using the angular variable $\Delta\phi(j_1,\MET)$. The ratio variables alone are able to exploit the ISR modified kinematics.  Indeed, much harder cuts on ratio variables are derived compared to the previous Tight-$p_T$ and Loose-$p_T$ cuts at the 8 TeV LHC. This shows the usefulness of the ratio variables especially at the high luminosity stage. Table.~\ref{tab:300280} shows the corresponding cut-flow and \Fig{fig:reach14} shows the exclusion reach. In particular in \Fig{fig:reach14}, we show the results only for the small-gap region because the large-gap search is already well developed by the experimental collaborations. Again, we note that the $\Delta \lesssim 12$ GeV region is limited by the condition min$(\mSFOS)>12$ GeV. We expect that gauginos up to about 320 GeV (700 GeV) with $\Delta \sim 30$ GeV and 220 GeV (620 GeV) with $\Delta \sim 20$ GeV can be probed if the gauginos decay via gauge bosons (light on-shell sleptons).

\begin{table}[t] \centering
\begin{tabular}{c|l|c|c|c|c}
\hline \hline
300-280  & cuts & $S $ &  $\frac{S}{B}$ &  $\frac{S}{\sqrt{B}} $  &  $\frac{S}{\sqrt{B + (0.15\cdot B)^2}}$  \\
\hline \hline
Loose-$p_T$ baseline & ($p_T(\ell)>$7, $p_T(j)>$30,  & 56 & 0.018 & 1.0  & 0.12 \\
& min$(\mSFOS)>$12,  & & & &  \\
& $\mSFOS(Z)<$81) & & & & \\
\hline
& min$(\mSFOS)< \Delta$ & 50  & 0.049 & 1.6 & 0.32 \\
Loose-$p_T$ & $\MET>60,\, p_T(\ell_1)<50$   & 32 & 0.21 & 2.6 & 0.78 \\
(14) cuts& $p_T(\ell_1)/p_T(j_1)<0.2$& 17 & 0.64 & 3.3 & 2.59 \\
& $\MET/p_T(j_1) > 0.9$ & 13  & 1.2 & 3.9 & 3.44 \\
\hline \hline
\end{tabular}
\caption{Cut-flow with Loose-$p_T$(14) cuts for the $300-280$ benchmark at LHC14 300/fb. 
 The other details of the table are as described in Table~\ref{tab:150120}.}
\label{tab:300280}\end{table}

%%%%%%%%%%%%%%%%%%
%%%%%%%%%%%%%%%%%%
\section{Conclusions}\label{sec:conclusions}
 
In this paper, we propose new search strategies to enhance the sensitivity of the LHC electroweakino searches in  the small mass-gap region with $\Delta\equiv m(\chi_{{\rm{NLSP}}})-m(\chi_{{\rm{LSP}}})\sim(10-50)$ GeV. In particular, we focus on the $3\ell+\MET$ final state arising from the associated production of a chargino and a second lightest neutralino NLSPs. 
The main background for this channel is the $WZ$ productions.
 
One useful variable that distinguishes the signal from the background is the minimum value of the invariant masses of same sign opposite flavor leptons (min($\mSFOS)$). Signal events show a sharp edge at around $\Delta$ while the distribution of background events peak at around $m_Z$. Therefore vetoing events with min($\mSFOS)>\Delta$ can sizably increase the S/B ratio.
 
Additionally, we point out the relevance of a relatively energetic ISR jet. Specifically, we observe that the correlation between the missing energy and the $p_T$ of the ISR jet can be significantly different in the signal and background samples. The strong correlation between ${E}_T^{\rm miss}$ and $p_T(j_1)$ of the signal is reflected in the distribution of the ratio variable $\frac{\MET}{p_T(j_1)}$ which peaks at around 1. Furthermore, $\MET$ and $j_1$ are back-to-back direction in the transverse plane creating a large $\Delta \phi(\MET,j_1)$. These features are particularly interesting because they hold generically true in small gap scenarios independently on the overall SUSY mass scales. On the other hand, for the background, the correlation is weaker, and both $\frac{\MET}{p_T(j_1)}$ and $\Delta \phi(\MET, j_1)$ tend to be smaller.  
Moreover, to further increase the LHC sensitivity, we note that one can veto boosted leptons. Signal leptons from small gap scenarios are in fact soft and receive only a small boost from the $p_T$ of the ISR jet. Background leptons are instead harder and more sensitive to the boost of the ISR jet. This observation will be particularly relevant at the high luminosity stage, at which one could require a larger ISR boost. 

We note that triggering on very soft leptons may be an issue at the LHC. However, we expect that the LHC lepton triggers can be supplemented by having a relatively energetic ISR jet.
Another limitation of our search strategy arises from the requirement min$(\mSFOS)\gtrsim$ 12 GeV which is used to suppress the background coming from low mass hadronic resonances decaying to leptons. It would be interesting to develop techniques to manage and reduce this background which will eventually allows to relax the min($\mSFOS$) cut.

Using the proposed search strategies, we show that the LHC could already probe NLSP gauginos up to 130 (300) GeV if the mass splitting is $\Delta\sim 20$ GeV and gauginos decay via gauge bosons (light sleptons). The  14 TeV LHC reach is instead pushed to the multi-hundred GeV scale -- gauginos up to 220 (620) GeV with $\Delta\sim 20$ GeV and up to 320 (700) GeV with $\Delta\sim 30$ GeV can be probed with 300/fb data.

We finally note that our methods are based on general considerations on the interplay between the kinematics of the ISR and the decay products in the electroweakino cascade decays.
Therefore, we expect our methods can bring a similar enhancement of the sensitivity in complementary channels such as the same sign dilepton and single lepton channels.

%%%%%%%%%%%%%%%%%%
%%%%%%%%%%%%%%%%%%
\acknowledgments

 S.G. and L.T.W would like to thank KITP
for its hospitality during some part of this work has been carried out. S.J. is grateful to Chicago 2012 LHC workshop where part of this work is carried out. L.T.W. is supported by  the DOE Early Career Award under grant de-sc0003930. This work is
supported in part by DOE under Contract No. DE-AC02-06CH11357 (ANL), DE-FGO2-96-
ER40956 (U.Chicago) and by the National Science Foundation under Grant No. NSF PHY11-25915 (KITP).

%\paragraph{Note added.} 

%%%%%%%%%%%%%%%%%%
%%%%%%%%%%%%%%%%%%
\appendix 
\section{Validation of the Monte Carlo event samples} \label{sec:app-mc}

We validate our Monte Carlo (MC) $WZ$ background sample by comparing the $\mSFOS$ spectrum calculated using Madgraph events with that obtained from the NLO theoretical calculation in Ref.~\cite{Campbell:1999ah}. It is especially important to validate the $\mSFOS$ variable since, as we explained in Sec.~\ref{sec:SignalBackground}, it is one of the most useful variables discriminating between signal and background. Reproducing properly this spectrum may not be straightforward as it can be subject to various subtle effects: low-mass resonances, singular photonic backgrounds, and photon radiations off leptons. Thus, we specifically use this spectrum to test and validate our MC generation of the background. 

\begin{figure}[t] \centering
\includegraphics[width=0.48\textwidth]{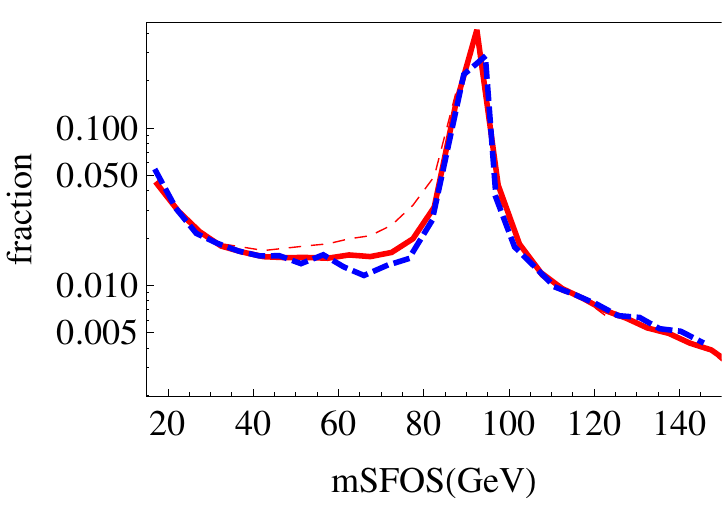}
\caption{Validation of our $WZ$ MC sample. The predictions of the invariant mass of SFOS leptons are compared at the 2 TeV Tevatron. Blue-dashed line: NLO theoretical calculation~\cite{Campbell:1999ah}, Red-dashed line: our results without photons clustered into nearby leptons, Red-thick line: our results after merging photons into nearby leptons. The events shown are satisfying the cuts described in Ref.~\cite{Campbell:1999ah}. If an event has two SFOS pairs, both invariant masses are histogramed with half the event weight. For comparison purposes, no lepton identification efficiency is applied. }
\label{fig:valid}
\end{figure}

It turns out that the photon radiation off leptons have a rather important effect in diluting the lepton momentum and modifying the invariant mass spectrum. The red dashed line in \Fig{fig:valid} shows that, ignoring the photons in the final state significantly smears the $Z$ boson peak toward low mass regions. 
The effect is particularly relevant for electrons, since they generically radiate more than muons. Not corrected properly, this effect would degrade the sensitivity of our analysis to small gap regions, since the $Z$ boson peak will overlap more with the low invariant mass of the signal. 

In practice, soft collinear photons radiated off of leptons will not be separately identified at the LHC. 
To keep this into account, we cluster photons into nearby leptons using the anti-$k_T$ algorithm with $R=0.4$. If the clustered object has a lepton inside, it is treated as a lepton. Furthermore, a lepton is called isolated if the scalar sum of the $p_T$ of all particles within a radius R=0.4 from the lepton is smaller than 0.2 times the $p_T$ of the lepton.
 As it can been seen from \Fig{fig:valid}, with this procedure, we can reproduce the NLO theoretical spectrum (blue dashed line in the figure) reasonably well. This test can not only validate our MC method, but also support the robustness of our computation of the $\mSFOS$ spectrum under higher order QCD corrections.

For a better comparison with Ref.~\cite{Campbell:1999ah}, in \Fig{fig:valid}, we do not apply any lepton identification efficiency. However, in our work, we cannot neglect this effect because we use soft leptons which suffer from relatively low identification efficiency.  
In particular, as shown in Table~\ref{tab:lepid}, for electrons we assume a $p_T$-dependent efficiency that drops from above $\sim 90\%$ to below 70\% for ${\cal O}(10)$ GeV~\cite{Aad:2011mk}. For muons instead, we use a flat efficiency of $91\%$ for $p_T(\mu)>7$ GeV \cite{Aad:2009wy}. The $\eta$-dependence is ignored.

Taking into account these effects, we are able to reproduce well the $WZ$ background results reported by the ATLAS collaboration in Ref.~\cite{ATLAS:2013rla}.

\begin{table}[t] \centering
\begin{tabular}{c||c|c|c|c|c|c}
\hline \hline
$p_T(e)$ (GeV)     & $(<15)$      & $(15-21)$       &  $(21-25)$        & $(25-30)$       & $(30-35)$        &  $>35$ \\
\hline
electron ID efficiency          & $66\%$     & $74\%$          &  $86\%$           & $90\%$          & $94\%$           &  $95\%$ \\
\hline \hline
\end{tabular}
\caption{$p_T$-dependent electron identification efficiency taken from Ref.~\cite{Aad:2011mk}. }
\label{tab:lepid}\end{table}

%%%%%%%%%%%%%%%%%%
%%%%%%%%%%%%%%%%%%
\section{Heavy versus light sleptons} \label{sec:app-slep}

In this section, we compare the distributions for the several kinematic variables presented in the text, in the case of heavy and light sleptons.
Differences may arise since, in the case of on-shell sleptons the chargino/second lightest neutralino decays proceed through two subsequent two-body decays. In the case of heavy sleptons, instead, the decays are three-body decays mediated by gauge bosons.

\begin{figure}[t] \centering
\includegraphics[width=\textwidth]{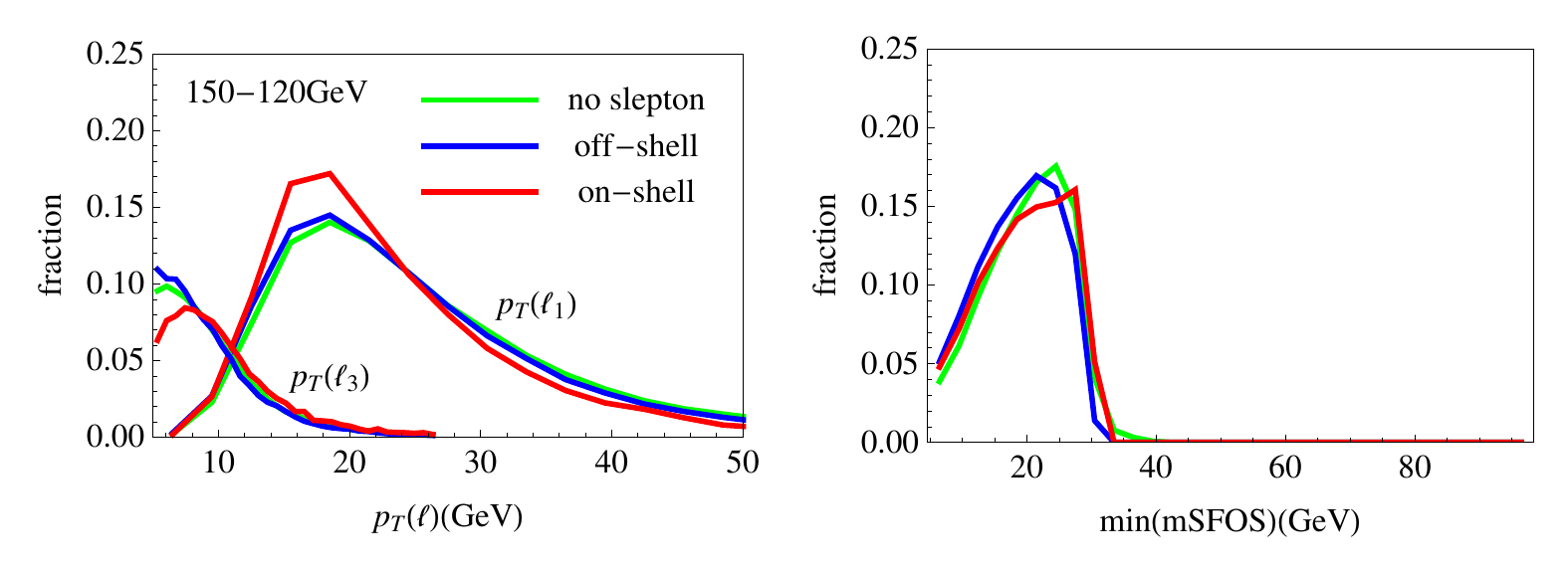}
\caption{Comparison between the lepton spectrums obtained from NLSP gauginos decaying via gauge bosons (``no slepton'' case with green curve), off-shell sleptons (blue curve), or on-shell sleptons (red curve). Events correspond to the (150-120) benchmark. }
\label{fig:onoffnoslep}
\end{figure}

In \Fig{fig:onoffnoslep}, we report the distributions for the variables that show the largest difference in the case of light/heavy sleptons: the $p_T$ of the hardest and of the softest lepton and min($\mSFOS)$. We use events arising from the (150-120) benchmark, but we note that the conclusion would not change, on condition of having a small-gap scenario. As we observe from the figure, the differences between the case of on-shell sleptons (in red) and heavy sleptons (in green) are minimal. Furthermore, in \Fig{fig:onoffnoslep}, we also show the spectrums coming from a scenario with light, but off-shell, sleptons (we fix $m_{\tilde\ell_L}=m_{\tilde\nu}=200$ GeV for the benchmark represented in blue in the figure).  Again we note that the distributions differ only slightly.

We finally note that, in more realistic SUSY spectra, sleptons and sneutrinos are split in mass. We checked that the distributions for our kinematic variables do not appreciably change introducing a splitting between sleptons and sneutrinos. 

For all these reasons, we apply our cuts optimized for the case of heavy sleptons to the case of light sleptons (see Sec.~\ref{sec:results})

%%%%%%%%%%%%%%%%%%
%%%%%%%%%%%%%%%%%%
\bibliography{Bibliography}

\providecommand{\href}[2]{#2}\begingroup\raggedright\begin{thebibliography}{10}

\bibitem{ATLAS:2013sma}
{\bf ATLAS} Collaboration, {\it {Search for squarks and gluinos using final
  states with jets and missing transverse momentum with the ATLAS detector in
  $\sqrt{s}$ = 8 TeV proton-proton collisions}}, . ATLAS-CONF-2013-047.

\bibitem{ATLAS:2013062}
{\bf ATLAS} Collaboration, {\it {Search for squarks and gluinos in events with
  isolated leptons, jets and missing transverse momentum at $\sqrt{s}$=8 TeV
  with the ATLAS detector}}, . ATLAS-CONF-2013-062.

\bibitem{ATLAS:2013tma}
{\bf ATLAS} Collaboration, {\it {Search for strongly produced superpartners in
  final states with two same sign leptons with the ATLAS detector using 21 fb-1
  of proton-proton collisions at sqrt(s)=8 TeV.}}, . ATLAS-CONF-2013-007.

\bibitem{Chatrchyan:2013lya}
{\bf CMS} Collaboration, S.~Chatrchyan {\em et.~al.}, {\it {Search for
  supersymmetry in hadronic final states with missing transverse energy using
  the variables $\alpha_T$ and b-quark multiplicity in pp collisions at
  $\sqrt{s}$ = 8 TeV}},  \href{http://xxx.lanl.gov/abs/1303.2985}{{\tt
  arXiv:1303.2985}}.

\bibitem{CMS:2012yua}
{\bf CMS} Collaboration, {\it {Search for supersymmetry with the razor
  variables at CMS}}, . CMS-PAS-SUS-12-005.

\bibitem{ArkaniHamed:2004fb}
N.~Arkani-Hamed and S.~Dimopoulos, {\it {Supersymmetric unification without low
  energy supersymmetry and signatures for fine-tuning at the LHC}},  {\em JHEP}
  {\bf 0506} (2005) 073, [\href{http://xxx.lanl.gov/abs/hep-th/0405159}{{\tt
  hep-th/0405159}}].

\bibitem{Giudice:2004tc}
G.~Giudice and A.~Romanino, {\it {Split supersymmetry}},  {\em Nucl.Phys.} {\bf
  B699} (2004) 65--89, [\href{http://xxx.lanl.gov/abs/hep-ph/0406088}{{\tt
  hep-ph/0406088}}].

\bibitem{Randall:1998uk}
L.~Randall and R.~Sundrum, {\it {Out of this world supersymmetry breaking}},
  {\em Nucl.Phys.} {\bf B557} (1999) 79--118,
  [\href{http://xxx.lanl.gov/abs/hep-th/9810155}{{\tt hep-th/9810155}}].

\bibitem{Giudice:1998xp}
G.~F. Giudice, M.~A. Luty, H.~Murayama, and R.~Rattazzi, {\it {Gaugino mass
  without singlets}},  {\em JHEP} {\bf 9812} (1998) 027,
  [\href{http://xxx.lanl.gov/abs/hep-ph/9810442}{{\tt hep-ph/9810442}}].

\bibitem{Giudice:2010wb}
G.~F. Giudice, T.~Han, K.~Wang, and L.-T. Wang, {\it {Nearly Degenerate
  Gauginos and Dark Matter at the LHC}},  {\em Phys.Rev.} {\bf D81} (2010)
  115011, [\href{http://xxx.lanl.gov/abs/1004.4902}{{\tt arXiv:1004.4902}}].

\bibitem{ATLAS:2013rla}
{\bf ATLAS} Collaboration, {\it {Search for direct production of charginos and
  neutralinos in events with three leptons and missing transverse momentum in
  21$\,$fb$^{-1}$ of pp collisions at $\sqrt{s}=8\,$TeV with the ATLAS
  detector}}, . ATLAS-CONF-2013-035.

\bibitem{Chatrchyan:2013006}
{\bf CMS} Collaboration, {\it {Search for electroweak production of charginos,
  neutralinos and sleptons using leptonic final states in $pp$ collisions at
  $\sqrt{s}=8$ TeV}}, . CMS-SUS-13-006.

\bibitem{ATLAS:2013zka}
{\bf ATLAS} Collaboration, {\it {Search for direct-slepton and direct-chargino
  production in final states with two opposite-sign leptons, missing transverse
  momentum and no jets in 20/fb of pp collisions at $\sqrt{s}$ = 8 TeV with the
  ATLAS detector}}, . ATLAS-CONF-2013-049.

\bibitem{Aad:2012jva}
{\bf ATLAS} Collaboration, G.~Aad {\em et.~al.}, {\it {Search for supersymmetry
  in events with photons, bottom quarks, and missing transverse momentum in
  proton-proton collisions at a centre-of-mass energy of 7 TeV with the ATLAS
  detector}},  {\em Phys.Lett.} {\bf B719} (2013) 261--279,
  [\href{http://xxx.lanl.gov/abs/1211.1167}{{\tt arXiv:1211.1167}}].

\bibitem{Chatrchyan:2012bba}
{\bf CMS Collaboration} Collaboration, S.~Chatrchyan {\em et.~al.}, {\it
  {Search for new physics in events with photons, jets, and missing transverse
  energy in $pp$ collisions at $\sqrt{s}=7$ TeV}},  {\em JHEP} {\bf 1303}
  (2013) 111, [\href{http://xxx.lanl.gov/abs/1211.4784}{{\tt
  arXiv:1211.4784}}].

\bibitem{Meade:2009qv}
P.~Meade, M.~Reece, and D.~Shih, {\it {Prompt Decays of General Neutralino
  NLSPs at the Tevatron}},  {\em JHEP} {\bf 1005} (2010) 105,
  [\href{http://xxx.lanl.gov/abs/0911.4130}{{\tt arXiv:0911.4130}}].

\bibitem{Gunion:2001fu}
J.~F. Gunion and S.~Mrenna, {\it {Probing models with near degeneracy of the
  chargino and LSP at a linear e+ e- collider}},  {\em Phys.Rev.} {\bf D64}
  (2001) 075002, [\href{http://xxx.lanl.gov/abs/hep-ph/0103167}{{\tt
  hep-ph/0103167}}].

\bibitem{Carena:2008mj}
M.~Carena, A.~Freitas, and C.~Wagner, {\it {Light Stop Searches at the LHC in
  Events with One Hard Photon or Jet and Missing Energy}},  {\em JHEP} {\bf
  0810} (2008) 109, [\href{http://xxx.lanl.gov/abs/0808.2298}{{\tt
  arXiv:0808.2298}}].

\bibitem{Drees:2012dd}
M.~Drees, M.~Hanussek, and J.~S. Kim, {\it {Light Stop Searches at the LHC with
  Monojet Events}},  {\em Phys.Rev.} {\bf D86} (2012) 035024,
  [\href{http://xxx.lanl.gov/abs/1201.5714}{{\tt arXiv:1201.5714}}].

\bibitem{Alvarez:2012wf}
E.~Alvarez and Y.~Bai, {\it {Reach the Bottom Line of the Sbottom Search}},
  {\em JHEP} {\bf 1208} (2012) 003,
  [\href{http://xxx.lanl.gov/abs/1204.5182}{{\tt arXiv:1204.5182}}].

\bibitem{Alves:2012ft}
D.~S. Alves, M.~R. Buckley, P.~J. Fox, J.~D. Lykken, and C.-T. Yu, {\it {Stops
  and MET: The Shape of Things to Come}},  {\em Phys.Rev.} {\bf D87} (2013)
  035016, [\href{http://xxx.lanl.gov/abs/1205.5805}{{\tt arXiv:1205.5805}}].

\bibitem{Bartels:2012ex}
C.~Bartels, M.~Berggren, and J.~List, {\it {Characterising WIMPs at a future
  $e^+e^-$ Linear Collider}},  {\em Eur.Phys.J.} {\bf C72} (2012) 2213,
  [\href{http://xxx.lanl.gov/abs/1206.6639}{{\tt arXiv:1206.6639}}].

\bibitem{Dreiner:2012gx}
H.~K. Dreiner, M.~Kramer, and J.~Tattersall, {\it {How low can SUSY go?
  Matching, monojets and compressed spectra}},  {\em Europhys.Lett.} {\bf 99}
  (2012) 61001, [\href{http://xxx.lanl.gov/abs/1207.1613}{{\tt
  arXiv:1207.1613}}].

\bibitem{Delgado:2012eu}
A.~Delgado, G.~F. Giudice, G.~Isidori, M.~Pierini, and A.~Strumia, {\it {The
  light stop window}},  {\em Eur.Phys.J.} {\bf C73} (2013) 2370,
  [\href{http://xxx.lanl.gov/abs/1212.6847}{{\tt arXiv:1212.6847}}].

\bibitem{Berggren:2013vfa}
M.~Berggren, F.~Brummer, J.~List, G.~Moortgat-Pick, T.~Robens, {\em et.~al.},
  {\it {Tackling light higgsinos at the ILC}},
  \href{http://xxx.lanl.gov/abs/1307.3566}{{\tt arXiv:1307.3566}}.

\bibitem{Abreu:2000as}
{\bf DELPHI} Collaboration, P.~Abreu {\em et.~al.}, {\it {Update of the search
  for charginos nearly mass-degenerate with the lightest neutralino}},  {\em
  Phys.Lett.} {\bf B485} (2000) 95--106,
  [\href{http://xxx.lanl.gov/abs/hep-ex/0103035}{{\tt hep-ex/0103035}}].

\bibitem{Heister:2002mn}
{\bf ALEPH} Collaboration, A.~Heister {\em et.~al.}, {\it {Search for charginos
  nearly mass degenerate with the lightest neutralino in e+ e- collisions at
  center-of-mass energies up to 209-GeV}},  {\em Phys.Lett.} {\bf B533} (2002)
  223--236, [\href{http://xxx.lanl.gov/abs/hep-ex/0203020}{{\tt
  hep-ex/0203020}}].

\bibitem{Abbiendi:2002vz}
{\bf OPAL} Collaboration, G.~Abbiendi {\em et.~al.}, {\it {Search for nearly
  mass degenerate charginos and neutralinos at LEP}},  {\em Eur.Phys.J.} {\bf
  C29} (2003) 479--489, [\href{http://xxx.lanl.gov/abs/hep-ex/0210043}{{\tt
  hep-ex/0210043}}].

\bibitem{Alwall:2011uj}
J.~Alwall, M.~Herquet, F.~Maltoni, O.~Mattelaer, and T.~Stelzer, {\it {MadGraph
  5 : Going Beyond}},  {\em JHEP} {\bf 1106} (2011) 128,
  [\href{http://xxx.lanl.gov/abs/1106.0522}{{\tt arXiv:1106.0522}}].

\bibitem{Sjostrand:2006za}
T.~Sjostrand, S.~Mrenna, and P.~Z. Skands, {\it {PYTHIA 6.4 Physics and
  Manual}},  {\em JHEP} {\bf 0605} (2006) 026,
  [\href{http://xxx.lanl.gov/abs/hep-ph/0603175}{{\tt hep-ph/0603175}}].

\bibitem{Mangano:2006rw}
M.~L. Mangano, M.~Moretti, F.~Piccinini, and M.~Treccani, {\it {Matching matrix
  elements and shower evolution for top-quark production in hadronic
  collisions}},  {\em JHEP} {\bf 0701} (2007) 013,
  [\href{http://xxx.lanl.gov/abs/hep-ph/0611129}{{\tt hep-ph/0611129}}].

\bibitem{Cacciari:2008gp}
M.~Cacciari, G.~P. Salam, and G.~Soyez, {\it {The Anti-k(t) jet clustering
  algorithm}},  {\em JHEP} {\bf 0804} (2008) 063,
  [\href{http://xxx.lanl.gov/abs/0802.1189}{{\tt arXiv:0802.1189}}].

\bibitem{Cacciari:2011ma}
M.~Cacciari, G.~P. Salam, and G.~Soyez, {\it {FastJet User Manual}},  {\em
  Eur.Phys.J.} {\bf C72} (2012) 1896,
  [\href{http://xxx.lanl.gov/abs/1111.6097}{{\tt arXiv:1111.6097}}].

\bibitem{Beenakker:1996ed}
W.~Beenakker, R.~Hopker, and M.~Spira, {\it {PROSPINO: A Program for the
  production of supersymmetric particles in next-to-leading order QCD}},
  \href{http://xxx.lanl.gov/abs/hep-ph/9611232}{{\tt hep-ph/9611232}}.

\bibitem{Campbell:2010ff}
J.~M. Campbell and R.~Ellis, {\it {MCFM for the Tevatron and the LHC}},  {\em
  Nucl.Phys.Proc.Suppl.} {\bf 205-206} (2010) 10--15,
  [\href{http://xxx.lanl.gov/abs/1007.3492}{{\tt arXiv:1007.3492}}].

\bibitem{Aad:2011mk}
{\bf ATLAS} Collaboration, G.~Aad {\em et.~al.}, {\it {Electron performance
  measurements with the ATLAS detector using the 2010 LHC proton-proton
  collision data}},  {\em Eur.Phys.J.} {\bf C72} (2012) 1909,
  [\href{http://xxx.lanl.gov/abs/1110.3174}{{\tt arXiv:1110.3174}}].

\bibitem{Aad:2009wy}
{\bf ATLAS} Collaboration, G.~Aad {\em et.~al.}, {\it {Expected Performance of
  the ATLAS Experiment - Detector, Trigger and Physics}},
  \href{http://xxx.lanl.gov/abs/0901.0512}{{\tt arXiv:0901.0512}}. SLAC-R-980,
  CERN-OPEN-2008-020.

\bibitem{Rolbiecki:2012gn}
K.~Rolbiecki and K.~Sakurai, {\it {Constraining compressed supersymmetry using
  leptonic signatures}},  {\em JHEP} {\bf 1210} (2012) 071,
  [\href{http://xxx.lanl.gov/abs/1206.6767}{{\tt arXiv:1206.6767}}].

\bibitem{Sfyrla:2013yva}
{\bf ATLAS Collaboration} Collaboration, A.~Sfyrla, {\it {ATLAS triggering on
  SUSY in 2012}},  {\em EPJ Web Conf.} {\bf 49} (2013) 18014.

\bibitem{Zarzhitsky:2008zz}
P.~Zarzhitsky, {\it {Search for supersymmetry in a three lepton plus jets
  signature}}, . AAT-3316299, PROQUEST-1588787421.

\bibitem{Campbell:1999ah}
J.~M. Campbell and R.~K. Ellis, {\it {An Update on vector boson pair production
  at hadron colliders}},  {\em Phys.Rev.} {\bf D60} (1999) 113006,
  [\href{http://xxx.lanl.gov/abs/hep-ph/9905386}{{\tt hep-ph/9905386}}].

\end{thebibliography}\endgroup
\bibliographystyle{JHEP}

\end{document}